\documentclass[aps,twocolumn,superscriptaddress,showpacs]{revtex4}

\usepackage[dvips]{graphicx}
\newcommand{\nc}{\newcommand}           
\nc{\vc}[1]     {\mbox{\boldmath $#1$}} 
\nc{\mapleft}[1]{                       
 \smash{\mathop{                      %
  \hbox to 0.90cm{\rightarrowfill} }\limits_{#1}}}

\nc{\mydraft}	{\setlength{\topmargin}{-1.5cm}}
\mydraft

\begin{document}

\title{Four-body resonances of $^7$B using the complex scaling method}

\author{Takayuki Myo\footnote{myo@ge.oit.ac.jp}}
\affiliation{General Education, Faculty of Engineering, Osaka Institute of Technology, Osaka, Osaka 535-8585, Japan}
\affiliation{Research Center for Nuclear Physics (RCNP), Osaka University, Ibaraki 567-0047, Japan}

\author{Yuma Kikuchi\footnote{yuma@rcnp.osaka-u.ac.jp}}
\affiliation{Research Center for Nuclear Physics (RCNP), Osaka University, Ibaraki 567-0047, Japan}

\author{Kiyoshi Kat\=o\footnote{kato@nucl.sci.hokudai.ac.jp}}
\affiliation{Division of Physics, Graduate School of Science, Hokkaido University, Sapporo 060-0810, Japan}

\date{\today}

\begin{abstract}
We study the resonance spectroscopy of the proton-rich nucleus $^7$B in the $^4$He+$p$+$p$+$p$ cluster model.
Many-body resonances are treated on the correct boundary condition as the Gamow states using the complex scaling method.
We predict five resonances of $^7$B and evaluate the spectroscopic factors of the $^6$Be-$p$ components.
The importance of the $^6$Be($2^+$)-$p$ component is shown in several states of $^7$B, which is a common feature of $^7$He, a mirror nucleus of $^7$B.
For only the ground state of $^7$B, the mixing of $^6$Be($2^+$) state is larger than that of $^6$He($2^+$) in $^7$He, which indicates the breaking of the mirror symmetry.
This is caused by the small energy difference between $^7$B and the excited $^6$Be($2^+$) state, whose origin is the Coulomb repulsion.
\end{abstract}



\maketitle 

\section{Introduction}

The radioactive beam experiments have provided us with much information on unstable nuclei far from the stability.
In particular, the light nuclei near the drip-line exhibit new phenomena of nuclear structures,
such as the neutron halo structure found in $^6$He, $^{11}$Li and $^{11}$Be \cite{tanihata85}.
The unstable nuclei can often be unbound states beyond the particle thresholds due to the weak binding nature.
The resonance spectroscopy of unbound states beyond the drip-line has also been developed experimentally.
In addition to the energies and decay widths, the configuration properties are important to understand the structures of the resonances.
The spectroscopic factors ($S$-factors) give the useful information to know the configurations of extra nucleons in the resonances as well as in the weakly bound states.
It is also interesting to compare the structures of resonances and weakly bound states between proton-rich and neutron-rich sides, which is related to the mirror symmetry in unstable nuclei.

Recently, the experiment on $^7$B have been reported \cite{charity11} in addition to the old observation\cite{mcgrath67}.
The $^7$B nucleus is known as an unbound system beyond the proton drip-line and its ground state is naively considered to be the $3/2^-$ resonance.
The ground state of $^7$B is observed at 2 MeV above the $^6$Be+$p$ threshold energy and the excited states have never observed yet.
The $^7$B states can decay not only to two-body $^6$Be+$p$ channels, but also to many-body channels of $^5$Li+2$p$ and $^4$He+3$p$. This multi-particle decay condition makes difficulty to identify the states of $^7$B experimentally.
The mirror nucleus of $^7$B is $^7$He, which is also unbound system with respect to the one neutron emission.
Recent experiments of $^7$He have been reported \cite{Ko99,Bo01,Me02,Bo05,Sk06,Ry06,beck07,Wu08} 
and confirmed that its ground state is assigned to be the $3/2^-$ resonance.
The $S$-factor of $^6$He-$n$ component was reported for the ground state of $^7$He\cite{beck07}.
The excited states of $^7$He can decay into the $^4$He+3$n$ channel, 
which also makes difficulty to observe experimentally.
There still remain contradictions in the observed energy levels of $^7$He.

From the view point of the ``$^4$He+three protons / neutrons'' system,
the information of $^7$B and $^7$He is important to understand the structures outside the drip-lines as a four-body picture.
It is also interesting to examine the effect of Coulomb interaction and the mirror symmetry in the resonances of two nuclei.
Structures of resonances generally depend on the existence of the open channels as the thresholds of the particle emissions.
In this sense, the mirror symmetry of resonances can be related to the coupling behavior to the open channels.
It is interesting to compare the effects of the couplings to the open channels for the resonances of $^7$B and $^7$He.

In the theoretical side to treat the unbound states explicitly, several methods have been developed,
such as the microscopic cluster model \cite{adahchour06,arai09}, the continuum shell model \cite{volya05} and the Gamow shell model \cite{betan09,michel07}.
It is, however, difficult to satisfy the multiparticle decay conditions correctly for all open channels. 
For $^7$B, it is necessary to describe the $^4$He+3$p$ four-body resonances in the theory.
So far, no theory describes the $^7$B nucleus as four-body resonances.
It is also important to reproduce the threshold energies of subsystems for particle decays, namely, the positions of open channels.
Emphasizing these theoretical conditions, in this study, we employ the cluster orbital shell model (COSM) \cite{suzuki88,masui06,myo077,myo09} of the $^4$He+$3p$ four-body system.
In COSM, the effects of all open channels are taken into account explicitly\cite{myo077}, so that we can treat the many-body decaying phenomena.
In our previous works of neutron-rich systems\cite{myo077,myo09,myo10}, we have successfully described 
the He isotopes with the $^4$He+$4n$ model up to the five-body resonances of $^8$He including the full couplings with $^{5,6,7}$He.
We have described many-body resonances using the complex scaling method (CSM) \cite{ho83,moiseyev98,aoyama06} under the correct boundary conditions for all decay channels. 
In CSM, the resonant wave functions are directly obtained by diagonalization of the complex-scaled Hamiltonian using the $L^2$ basis functions.
The successful results of light nuclei using CSM have been obtained for energies, decay widths, spectroscopic factors and also for the breakup strengths induced by the Coulomb excitations\cite{myo01,myo0711}, 
monopole transition\cite{myo10} and one-neutron removal\cite{myo09}. 
Recently, CSM has been developed to apply to the nuclear reaction methods such as 
the scattering amplitude calculation \cite{kruppa07}, Lippmann-Schwinger equation\cite{kikuchi10} and the CDCC method\cite{matsumoto10}.

In this study, we proceed with our study of resonance spectroscopy to the proton-rich nucleus, $^7$B.
It is interesting to examine how our model describes $^7$B as four-body resonances. 
We predict the resonances of $^7$B and investigate their configuration properties.
We extract the $S$-factors of the $^6$Be-$p$ components for every $^7$B resonances.
The above $S$-factors are useful for understanding the coupling behavior between $^6$Be and the last proton.
For mirror nucleus, $^7$He, we have performed the same analysis of the $S$-factors of the $^6$He-$n$ components \cite{myo09}, in which the large mixing of the $^6$He($2^+$) state is confirmed.
From the viewpoint of the mirror symmetry, we compare the structures of $^7$B with those of $^7$He and discuss the effect of the Coulomb interaction on the mirror symmetry.
Since two nuclei are both unbound, the coupling effect of the open channels is discussed.

In Sec.~\ref{sec:model}, we explain the complex-scaled COSM wave function and the method of obtaining the $S$-factors using CSM.
In Sec.~\ref{sec:result}, we discuss the $^7$B structures and the $S$-factors of the $^6$Be-$p$ components.
Summary is given in Sec.~\ref{sec:summary}.

\section{Complex-scaled COSM}\label{sec:model}

\subsection{COSM for the $^4$He+\vc{N_{\rm v} p} systems}

We use COSM of the $^4$He+$N_{\rm v} p$ systems, where $N_{\rm v}$ is a valence proton number around $^4$He, 
namely, $N_{\rm v}=3$ for $^7$B.
The Hamiltonian form is the same as that used in Refs.~\cite{myo077,myo09};
\begin{eqnarray}
	H
&=&	\sum_{i=1}^{N_{\rm v}+1}{t_i} - T_G + \sum_{i=1}^{N_{\rm v}} V^{\alpha p}_i + \sum_{i<j}^{N_{\rm v}} V^{pp}_{ij}
    \\
&=&	\sum_{i=1}^{N_{\rm v}} \left[ \frac{\vec{p}^2_i}{2\mu} + V^{\alpha p}_i \right] + \sum_{i<j}^{N_{\rm v}} \left[ \frac{\vec{p}_i\cdot \vec{p}_j}{4m} + V^{pp}_{ij} \right] ,
    \label{eq:Ham}
\end{eqnarray}
where $t_i$ and $T_G$ are the kinetic energies of each particle ($p$ and $^4$He) and of the center of mass of the total system, respectively.
The operator $\vec{p}_i$ is the relative momentum between $p$ and $^4$He. 
The reduced mass $\mu$ is $4m/5$ using a nucleon mass $m$.
The $^4$He-$p$ interaction $V^{\alpha p}$ is given by the microscopic KKNN potential \cite{aoyama06,kanada79} for nuclear part,
in which the tensor correlation of $^4$He is renormalized on the basis of the resonating group method in the $^4$He+$N$ scattering.
For the Coulomb part, we use the folded Coulomb potential using the density of $^4$He having the $(0s)^4$ configuration.
We use the Minnesota potential \cite{tang78} as a nuclear part of $V^{pp}$ in addition to the Coulomb interaction.
These interactions reproduce the low-energy scattering of the $^4$He-$N$ and the $N$-$N$ systems, respectively.

For the wave function, $^4$He is treated as the $(0s)^4$ configuration of a harmonic oscillator wave function, 
whose length parameter is 1.4 fm to fit the charge radius of $^4$He as 1.68 fm.
The motion of valence protons around $^4$He is solved variationally using the few-body technique.
We expand the relative wave functions of the $^4$He+$N_{\rm v} p$ system using the COSM basis states \cite{suzuki88,masui06,myo077,myo09}.
In COSM, the total wave function $\Psi^J$ with a spin $J$ is represented by the superposition of the configuration $\Psi^J_c$ as
\begin{eqnarray}
    \Psi^J
&=& \sum_c C^J_c \Psi^J_c,
    \label{WF0}
    \\
    \Psi^J_c
&=& \prod_{i=1}^{N_{\rm v}} a^\dagger_{\alpha_i}|0\rangle, 
    \label{WF1}
\end{eqnarray}
where the vacuum $|0\rangle$ is given by the $^4$He ground state.
The creation operator $a^\dagger_{\alpha}$ is for the single particle state of a valence proton above $^4$He
with the quantum number $\alpha=\{n,\ell,j\}$ in a $jj$ coupling scheme.
Here, the index $n$ represents the different radial component. 
The coefficient $C^J_c$ represents the amplitude of the configuration 
and its index $c$ represents the set of $\alpha_i$ as $c=\{\alpha_1,\cdots,\alpha_{N_{\rm v}}\}$.
We take a summation over the available configurations in Eq.~(\ref{WF0}), which give a total spin $J$.

\begin{figure}[t]
\centering
\includegraphics[width=7.5cm,clip]{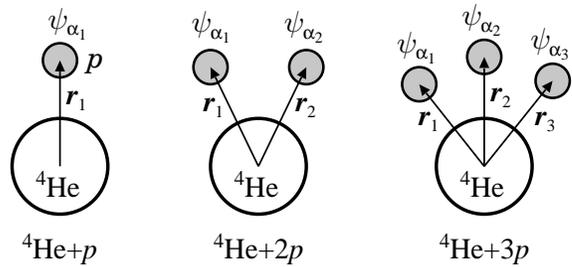}
\caption{Sets of the spatial coordinates in COSM for the $^4$He+$N_{\rm v} p$ system.}
\label{fig:COSM}
\end{figure}

The coordinate representation of the single particle state corresponding to $a^\dagger_{\alpha}$ is given as 
$\psi_{\alpha}(\vc{r})$ as function of the relative coordinate $\vc{r}$ between the center of mass of $^4$He 
and a valence proton \cite{suzuki88}, as shown in Fig.~\ref{fig:COSM}.
Considering the angular momentum coupling, the explicit wave functions of the COSM configuration $\Psi^J_c$ in Eq.~(\ref{WF1}) are expressed as
\begin{eqnarray}
    \Psi^J_c
&=& {\cal A}^\prime \left\{\, [\Phi(^4{\rm He}), \chi^{J}_c(N_{\rm v} p)]^J\, \right\},
    \label{eq:WF}
    \\
    \chi^{J}_c(p)
&=& \psi_{\alpha_1}^J,
    \\
    \chi^{J}_c(2p)
&=& {\cal A}\{ [\psi_{\alpha_1},\psi_{\alpha_2}]_J \},
    \label{eq:WF6}
    \\
    \chi^{J}_c(3p)
&=& {\cal A}\{ [[\psi_{\alpha_1},\psi_{\alpha_2}]_{j_{12}},\psi_{\alpha_3}]_J \}.
    \label{eq:WF7}
\end{eqnarray}
Here, $\Phi(^4{\rm He})$ is the $^4$He wave function with spin $0^+$.
The function $\chi^J_c(N_{\rm v} p)$ expresses the COSM wave functions for the valence protons.
The spin $j_{12}$ is a coupled angular momentum of the first and second valence protons. 
The antisymmetrizers between valence protons and
between a valence proton and nucleons in $^4$He are expressed as the symbols ${\cal A}$ and ${\cal A}^\prime$, respectively.
The effect of ${\cal A}^\prime$ is treated in the orthogonality condition model\cite{myo09,aoyama06}, in which
$\psi_{\alpha}$ is imposed to be orthogonal to the $0s$ state occupied by $^4$He.
We employ a sufficient number of radial bases of $\psi_\alpha$ to describe the spatial extension of valence protons in the resonances, in which $\psi_\alpha$ are normalized.
In this model, the radial part of $\psi_\alpha$ is expanded with the Gaussian basis functions for each orbit as
\begin{eqnarray}
    \psi_\alpha
&=& \sum_{k=1}^{N_{\ell j}} d^k_{\alpha}\ \phi_{\ell j}^k(\vc{r},b_{\ell j}^k),
    \label{WFR}
    \\
    \phi_{\ell j}^k(\vc{r},b_{\ell j}^k)
&=& {\cal N}\, r^{\ell} e^{-(r/b_{\ell j}^k)^2/2} [Y_{\ell}(\hat{\vc{r}}),\chi^\sigma_{1/2}]_{j}.
    \label{Gauss}
\end{eqnarray}
The index $k$ is for the Gaussian basis with the length parameter $b_{\ell j}^k$.
Normalization factor of the basis and a basis number are given by ${\cal N}$ and $N_{\ell j}$, respectively. 

In the COSM using Gaussian expansion, the total wave function $\Psi^J$ contains two-kinds of the expansion coefficients $\{C_c^J\}$ in Eq.~(\ref{WF0}) for configuration and $\{d^k_\alpha\}$ in Eq.~(\ref{WFR})
for each valence proton.
We determine them in the following procedure:
First, we solve the eigenvalue problem of the norm matrix of the Gaussian basis set in Eq.~(\ref{Gauss}), which are non-orthogonal, with the dimension $N_{\ell j}$. 
The coefficients $\{d^k_\alpha\}$ are determined to construct the orthonormalized single-particle basis set $\{\psi_\alpha\}$ 
having different radial components with the number $N_{\ell j}$.
Second, Hamiltonian matrix elements are constructed using $\{\psi_\alpha\}$ and diagonalized to determine $\{C_c^J\}$ from the variational principle.
The relation $\sum_{c} \left(C_c^J\right)^2=1$ is satisfied due to the normalization of the total wave function.
The same method of determining the expansion coefficients using Gaussian bases is used in the tensor-optimized shell model\cite{myo11}.

The numbers of the radial bases $N_{\ell j}$ of $\psi_\alpha$ are determined to converge the physical solutions $\Psi^J$.
The length parameters $b_{\ell j}^k$ are chosen in geometric progression \cite{myo09,aoyama06}.
We use at most 17 Gaussian basis functions by setting $b_{\ell j}^k$ from 0.2 fm to around 40 fm
with the geometric ratio of 1.4 as a typical one.
Due to the expansion of the radial wave function using a finite number of basis states, 
all the energy eigenvalues are discretized for bound, resonant and continuum states.
For reference, in the Gamow shell model calculation \cite{betan09,michel07}, 
the single particle states $\psi_\alpha$ consist of the resonant and the discretized continuum states obtained with the single particle potential
$V^{\alpha p}$ in Eq.~(\ref{eq:Ham}).

For $^7$B, all the channels of $^6$Be+$p$, $^5$Li+$2p$ and $^4$He+$3p$ are automatically included in the total COSM wave function $\Psi^J$.
These components are coupled to each other via the interactions and the antisymmetrization.
The couplings depend on the relative distances between $^4$He and a valence proton and between the valence protons.
We explain the coupling behavior between $^4$He and valence protons in COSM.
This is related to the boundary condition of the proton emission in $^7$B,
which is important when the resonant and continuum states are treated\cite{myo077,myo0711,myo05}. 
As an example, we consider the coupling between $^{7}$B and the $^6$Be+$p$ configurations. 
Asymptotically, when the last proton is located far away from $^6${Be}, namely, $\vc{r}_3\to\infty$ in Fig.~\ref{fig:COSM},
any coupling between $^6$Be and a last proton disappears, and $^6$Be becomes its isolated eigenstate of the Hamiltonian in Eq.~(\ref{eq:Ham})
with $N_{\rm v}=2$.

\begin{eqnarray}
        \Psi^J(^{7}{\rm B})
&=&     \sum_c C_c^J  {\cal A}^\prime\left\{ [ \Phi(^{4}\mbox{He}), \chi_c^J(3p) ]^J \right\}
        \label{asympt0}
        \\
&\mapleft{\vc{r}_3\to\infty}&
        \left[ \Psi^{J^\prime}_\nu(^{6}\mbox{Be}), \psi_{\alpha_3} \right]^{J},
        \label{asympt1}
        \\
        \Psi^{J^\prime}_\nu(^{6}\mbox{Be})
&=&     \sum_c C_{c,\nu}^{J^\prime} {\cal A}^\prime \left\{ [ \Phi(^{4}\mbox{He}), \chi_{c,\nu}^{J^\prime}(2p) ]^{J^\prime} \right\},
        \label{asympt2}
\end{eqnarray}
where the spin $J$ and $J'$ are for $^7$B and $^6$Be, respectively,
and the index $\nu$ indicates the eigenstate of $^6$He.
The mixing coefficients $\{C^{J^\prime}_{c,\nu}\}$ and the wave function $\chi^{J^\prime}_{c,\nu}(2p)$ in Eq.~(\ref{asympt2}) 
are those of the $^6$Be eigenstates.
Hence, the wave function $\chi^{J}_c(3p)$ in Eq.~(\ref{asympt0}) satisfies the following asymptotic forms
\begin{eqnarray}
    \sum_c C_c^J \chi^{J}_c(3p)
&\mapleft{\vc{r}_3\to\infty} & \left( \sum_c C_{c,\nu}^{J^\prime} \chi^{J^\prime}_{c,\nu}(2p)\right) \psi_{\alpha_3} .
    \label{asympt3}
\end{eqnarray}
This relation implies that the wave function of three valence protons of $^7$B 
is asymptotically decomposed into $^6$Be and a last proton. 
Equations~(\ref{asympt0})-(\ref{asympt3}) determine the boundary condition of COSM.
Contrastingly, when a last proton comes close to $^6$Be, the last proton dynamically couples to the $^6$Be eigenstates $\Psi_\nu^{J^\prime}$.
This coupling depends on the relative distance between $^6$Be and a last proton,
and changes the $^6$Be configurations from the isolated eigenstates of $^6$Be.
In COSM, the structure change of $^6$Be inside $^7$B is determined variationally to optimize the $^7$B eigenstates.
The same discussion is applied to the asymptotic conditions for the $^5$Li+$2p$ and $^4$He+$3p$ configurations.
Hence, the proton emissions can be handled with the correct boundary conditions in COSM.

We explain the parameters of the model space of COSM and the Hamiltonian which are determined in the previous analyses of He isotope\cite{myo077,myo09}.
For the single-particle states, we take the angular momenta $\ell\le 2$ to keep the accuracy of the converged energy within 0.3 MeV of $^6$He with the $^4$He+$n$+$n$ model in comparison with the full space calculation\cite{aoyama06}. 
In this model, we adjust the two-neutron separation energy of $^6$He($0^+$) to the experiment of 0.975 MeV 
by taking the 173.7 MeV of the repulsive strength of the Minnesota potential instead of the original value of 200 MeV.
The adjustment of the $NN$ interaction is originated from the pairing correlation between valence protons with higher angular momenta $\ell>2$ \cite{aoyama06}.
Hence, the present model reproduces the observed energies of $^{6}$He and is applied to the proton-rich nuclei in this analysis.

\subsection{Complex scaling method (CSM)}

We explain CSM, which describes resonances and nonresonant continuum states \cite{ho83,moiseyev98,aoyama06}.
Hereafter, we refer to the nonresonant continuum states as simply the continuum states.
In CSM, we transform the relative coordinates of the $^4$He+$N_{\rm v} p$ system, as $\vc{r}_i \to \vc{r}_i\, e^{i\theta}$
for $i=1,\cdots,N_{\rm v}$, where $\theta$ is a scaling angle.
The Hamiltonian in Eq.~(\ref{eq:Ham}) is transformed into the complex-scaled Hamiltonian $H_\theta$, and the corresponding complex-scaled Schr\"odinger equation is given as
\begin{eqnarray}
	H_\theta\Psi^J_\theta
&=&     E\Psi^J_\theta .
	\label{eq:eigen}
\end{eqnarray}
The eigenstates $\Psi^J_\theta$ are obtained by solving the eigenvalue problem of $H_\theta$ in Eq.~(\ref{eq:eigen}).
In CSM, we obtain all the energy eigenvalues $E$ of bound and unbound states on a complex energy plane, governed by the ABC theorem \cite{ABC}.
In this theorem, it is proved that the boundary condition of resonances is transformed to one of the damping behavior at the asymptotic region.
This condition makes it possible to use the same method of obtaining the bound states and resonances. 
For a finite value of $\theta$, every Riemann branch cut starting from the different thresholds is commonly rotated down by $2\theta$.
Hence, the continuum states such as $^6$Be+$p$ and $^5$Li+2$p$ channels in $^7$B are obtained on the branch cuts rotated by the $-2\theta$ 
from the corresponding thresholds \cite{myo077,myo09}.
On the contrary, bound states and resonances are obtainable independently of $\theta$.
We can identify the resonance poles with complex eigenvalues: $E=E_r-i\Gamma/2$, where $E_r$ and $\Gamma$ are the resonance energies and the decay widths, respectively. 
In the wave function, the $\theta$ dependence is included in the expansion coefficients in Eqs.~(\ref{WF0}) and (\ref{WFR}) as $\{C_c^J(\theta)\}$ and $\{d_\alpha^k(\theta)\}$, respectively. 
The value of the angle $\theta$ is determined to search for the stationary point of each resonance in a complex energy plane\cite{aoyama06,ho83,moiseyev98}.

The resonant state generally has a divergent behavior at asymptotic distance and then its norm is defined by a singular integral such as using the convergent factor method\cite{aoyama06,homma97,romo68}.
In CSM, on the other hand, resonances are precisely described as eigenstates expanded in terms of the $L^2$ basis functions.
The amplitudes of the resonances are finite and normalized as $\sum_{c} \left(C_c^J(\theta)\right)^2=1$.
The Hermitian product is not applied due to the bi-orthogonal relation \cite{ho83,moiseyev98,berggren68}.
The matrix elements of resonances are calculated using the amplitudes obtained in CSM.

In this study, we discretize the continuum states in terms of the basis expansion, as shown in the figures of energy eigenvalue distributions in Refs. \cite{myo01,myo09,aoyama06}.
The reliability of the continuum discretization in CSM has already been shown using the continuum level density\cite{suzuki05} and the phase shift analysis\cite{kruppa07}.

\subsection{Spectroscopic factor of $^7$B}

We explain the $S$-factors of the $^6$Be-$p$ components for $^7$B.
As was explained in the previous study \cite{myo09}, since the resonant states generally give complex matrix elements,
The $S$-factors of resonant states are not necessarily positive definite and defined by the squared matrix elements
using the bi-orthogonal property \cite{berggren68} as
\begin{eqnarray}
    S^{J,\nu}_{J',\nu'}
&=& \sum_\alpha S^{J,\nu}_{J',\nu',\alpha}\, ,
\\
    S^{J,\nu}_{J',\nu',\alpha}
&=& \frac{1}{2J+1} \langle \widetilde{\Phi}^{J'}_{\nu'}||a_\alpha ||\Psi^J_\nu \rangle^2\, ,
    \label{eq:S}
\end{eqnarray}
where the annihilation operator $a_\alpha$ is for single valence proton with the state $\alpha$. 
The spin $J$ and $J'$ are for $^7$B and $^6$Be, respectively.
The index $\nu$ ($\nu'$) indicates the eigenstate of $^7$B ($^6$Be).
The wave function $\Phi^{J'}_{\nu'}$ is for $^6$Be.
In this expression, the values of $S^{J,\nu}_{J',\nu'}$ are allowed to be complex.
In general, an imaginary part of the $S$-factors often becomes large relative to the real part for a resonance having a large decay width.
Recently, the Gamow shell model calculation also discuss the $S$-factors of resonances \cite{michel10}.

The sum rule value of $S$-factors, which includes resonance contributions of the final states, can be considered \cite{myo077}.
When we count all the $S$-factors not only of resonances but also of the continuum states in the final states,
the summed value of the $S$-factors is equal to the associated particle number, which is a real value
and does not contain any imaginary part, as similar to the transition strength calculation\cite{myo01,myo03}.
For $^7$B into the $^6$Be-$p$ decomposition, the summed value of the $S$-factor $S^{J,\nu}_{J',\nu'}$ in Eq.~(\ref{eq:S}) by taking all the $^6$Be states, is given as
\begin{eqnarray}
    \sum_{J',\nu'}\ S^{J,\nu}_{J',\nu'}
&=& \sum_{\alpha,m}\ 
    \langle \widetilde{\Psi}^{JM}_\nu|a^\dagger_{\alpha,m} a_{\alpha,m}| \Psi^{JM}_\nu \rangle
    \nonumber
    \\
&=& 3\ ,
    \label{eq:sf-sum}
\end{eqnarray} 
where we use the completeness relation of $^6$Be as
\begin{eqnarray}
    1
&=& \sum_{J',M'}\sum_{\nu'}\hspace*{-0.5cm}\int |\Phi^{J'M'}_{\nu'}\rangle \langle \widetilde{\Phi}^{J'M'}_{\nu'}|.
\end{eqnarray}
Here $M$ ($M'$) and $m$ are the $z$-components of the angular-momenta of the wave functions of $^7$B ($^6$Be) and 
of the creation and annihilation operators of the valence protons, respectively.
It is found that the summed value of the $S$-factors for the $^6$Be states becomes the valence proton number $N_{\rm v}$ of $^7$B.
This discussion of the $S$-factors is valid when the complex scaling is operated.
It is also shown that $S$-factors of the resonances are invariant with respect to the scaling angle $\theta$ \cite{myo09,homma97}.

The present $S$-factors can be used to obtain the strengths of the proton removal reaction from $^7$B into $^6$Be as a function of the energy of $^6$Be.
In the calculation, the $S$-factors not only of the resonances, but also of the many-body continuum states for $^7$B and $^6$Be are necessary.
The complex-scaled Green's function is also used to calculate the strength distribution \cite{myo09,myo01,myo98}.
In fact, for neutron-rich case, we have shown the one-neutron removal strength distributions from $^7$He into the $^6$He states using CSM\cite{myo09}.
The strength into the three-body scattering states of $^6$He as $^4$He+$n$+$n$ was successfully obtained by using the complex-scaled wave function of $^6$He. 
It was shown that the $^6$He($2^+$) resonance generates a sharp peak at around the resonance energy in the distribution.

In the numerical calculation, we express the radial part of the operator $a_\alpha$ in Eq.~(\ref{eq:S}) 
using the complete set expanded by 40 Gaussian basis functions with the maximum range of 100 fm for each orbit. 
This treatment is sufficient to converge the $S$-factor results.

\section{Results}\label{sec:result}

\subsection{Energy spectra of $^5$Li, $^6$Be and $^7$B}

\begin{figure}[b]
\centering
\includegraphics[width=8.5cm,clip]{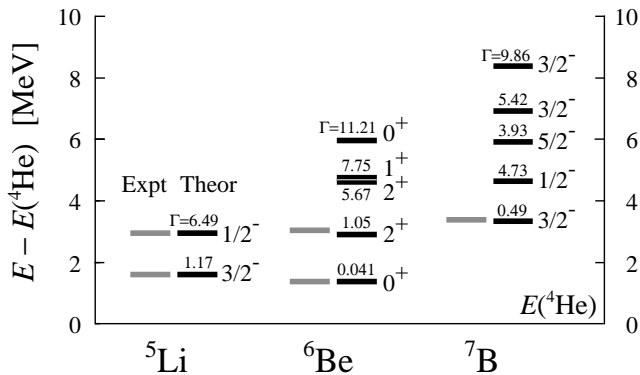}
\caption{Energy levels of $^5$Li, $^6$Be and $^7$B measured from the $^4$He energy. Units are in MeV.
Black and gray lines are theory and experiments, respectively. Small numbers are decay widths.}
\label{fig:B7}
\end{figure}

We show the systematic behavior of level structures of $^5$Li, $^6$Be and $^7$B in Fig. \ref{fig:B7}.
It is found that the present calculations agree with the observed energy levels. We furthermore predict many resonances for $^6$Be and $^7$B.
We first discuss the structures of $^6$Be, which are useful for the understanding of the $^7$B structures.
The $^6$Be states together with a last proton compose the thresholds of the decay of $^7$B. 
It is also interesting to compare the $^6$Be structures with those of $^6$He, a mirror and a neutron halo nucleus.

\begin{table}[t]
\caption{Energy eigenvalues of the $^{6}$Be resonances measured from the $^4$He+$p$+$p$ threshold.
The values with parentheses are the experimental ones\cite{aj89}. Dominant configurations are listed.}
\label{ene6}
\centering
\begin{ruledtabular}
\begin{tabular}{c|cccc}
          & Energy~[MeV]         &  Width~[MeV]      & Configuration        \\ \hline
 $0^+_1$  & $1.383$~($1.370$)    &  0.041(0.092)     & $(p_{3/2})^2$        \\
 $0^+_2$  & $5.95$               & 11.21             & $(p_{1/2})^2$        \\
 $2^+_1$  & $2.90$~($3.04$)       &  1.05 (1.16)      & $(p_{3/2})^2$        \\
 $2^+_2$  & $4.63$               &  5.67             & $(p_{3/2})(p_{1/2})$ \\
 $1^+$    & $4.76$               &  7.75             & $(p_{3/2})(p_{1/2})$ \\
\end{tabular}
\end{ruledtabular}
\end{table}

\begin{table}[t]
\caption{Components of the ground states of $^{6}$Be and $^6$He.}
\label{comp6_0}
\centering
\begin{ruledtabular}
\begin{tabular}{c|ccc}
Config.         & $^6$Be($0^+_1$) &  $^6$He($0^+_1$) \\ \hline
 $(p_{3/2})^2$  & $0.918-i0.006$  &  0.917  \\
 $(p_{1/2})^2$  & $0.041+i0.000$  &  0.043  \\
 $(1s_{1/2})^2$ & $0.010+i0.006$  &  0.009  \\
 $(d_{5/2})^2$  & $0.024+i0.000$  &  0.024  \\
 $(d_{3/2})^2$  & $0.007+i0.000$  &  0.007  \\
\end{tabular}
\end{ruledtabular}
\end{table}

\begin{table}[t]
\caption{Dominant components of the $2^+_1$ states of $^{6}$Be and $^6$He.}
\label{comp6_2}
\centering
\begin{ruledtabular}
\begin{tabular}{c|ccc}
Config.               & $^6$Be($2^+_1$) & $^6$He ($2^+_1$) \\ \hline
 $(p_{3/2})^2$        & $0.891+i0.030$  & $0.898+i0.013$ \\
 $(p_{3/2})(p_{1/2})$ & $0.097-i0.024$  & $0.089-i0.013$ \\
\end{tabular}
\end{ruledtabular}
\end{table}

\begin{table}[t]  
\caption{Radial properties of the ground states of $^6$Be and $^6$He in units of fm,
in comparison with the experiments of $^6$He; a\cite{tanihata92}, b\cite{alkazov97}, c\cite{kiselev05}, d\cite{mueller07}.}
\label{radius}
\centering
\begin{ruledtabular}
\begin{tabular}{c|cccc}
                 &  $^6$Be         & $^6$He          & $^6$He(exp.) \\ \hline
$R_{\rm m}$      &  2.80 + $i$0.17 &~~~2.37~~~       &  2.33(4)$^{\rm a}$,~2.30(7)$^{\rm b}$,~2.37(5)$^{\rm c}$   \\
$R_p$            &  3.13 + $i$0.20 & 1.82            &   \\
$R_n$            &  1.96 + $i$0.08 & 2.60            &   \\
$R_{\rm ch}$     &  3.25 + $i$0.21 & 2.01            &  2.068(11)$^{\rm d}$ \\
$r_{NN}$         &  6.06 + $i$0.35 & 4.82            &   \\
$r_{{\rm c}\mbox{-}2N}$ &  3.85 + $i$0.37 & 3.15     &   \\
$\theta_{NN}$    &  75.3           & 74.6            &   \\
\end{tabular}
\end{ruledtabular}
\end{table}

The resonance energies and the decay widths of $^6$Be are listed in Table \ref{ene6} with dominant configurations.
The components of each configuration for the $^6$Be and $^6$He ground states are listed in Table~\ref{comp6_0},
which are the square values of the amplitudes $\{C^J_c\}$ defined in Eq.~(\ref{WF0}).
We show the summation of the components belonging to the same configurations with different radial components of a valence proton.
It is noted that the amplitude of resonant wave function becomes a complex number and 
its real part can have a physical meaning when the imaginary part has relatively a small value.
It is confirmed that two ground states show the similar trend of configurations, which is dominated by $p$-shell. 
The configurations of the $2^+_1$ states of $^6$Be and $^6$He are also shown in Table~\ref{comp6_2},
where the energy and decay width of $^6$He($2^+_1$) are obtained as ($E_r$, $\Gamma$)=(0.879, 0.132) in MeV, measured from the $^4$He+$n$+$n$ threshold.
The good correspondence is seen for the dominant two configurations of the $2^+_1$ states.
These results indicate that the mirror symmetry is kept well for the configurations between $^6$Be and $^6$He.
Recently, Gamow shell model calculation discussed the $p$-shell contributions in the A=6 system\cite{michel10}.

The radial properties of $^6$Be are interesting to discuss the effect of the Coulomb repulsion in comparison with $^6$He having a halo structure, 
although the radius of $^6$Be can be complex numbers because of the resonance.
The results of the $^6$Be ground state are shown in Table \ref{radius} for matter ($R_{\rm m}$), proton ($R_p$), neutron($R_n$) charge ($R_{\rm ch}$) parts,
and the relative distances between valence nucleons ($r_{NN}$) and between the $^4$He core and the center of mass of two valence nucleons ($r_{{\rm c}\mbox{-}2N}$),
and the opening angle between two nucleons ($\theta_{NN}$) at the center of mass of the $^4$He core.
It is found that the values in $^6$Be are almost real, so that the real parts can be regarded to represent the radius properties of $^6$Be.
The distances between valence protons and between core and $2p$ in $^6$Be are wider than those of $^6$He by 26\% and 22\%, respectively. 
This result comes from the Coulomb repulsion between three constituents of $^4$He+$p$+$p$ in $^6$Be.
The Coulomb repulsion makes the energy of $^6$Be shift up to be a resonance in comparison with $^6$He, 
and also increases the relative distances between each constituent from the halo state of $^6$He.

\begin{figure}[t]
\centering
\includegraphics[width=8.0cm,clip]{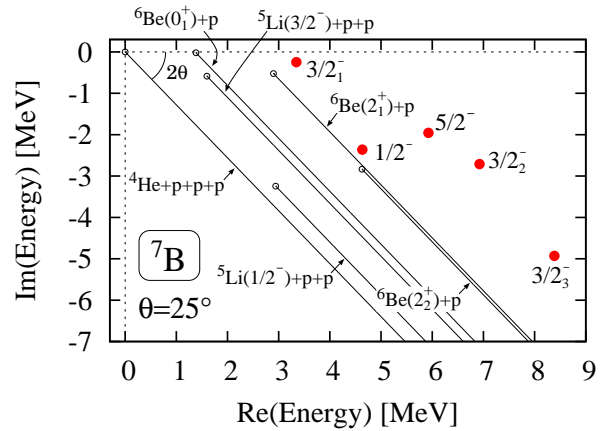}
\caption{(Color online) Energy eigenvalue distribution of $^7$B in complex energy plane.}
\label{fig:ene_7B}
\end{figure}

We discuss the structures of $^7$B.
The energy eigenvalues are listed in Table \ref{ene7} measured from the $^4$He+3$p$ threshold.
We obtain the five resonances which are all located above the $^6$Be($0^+_1$)+$p$ threshold, as shown in Fig.~\ref{fig:B7}, and four-body resonances.
In Fig.~\ref{fig:ene_7B}, we display the energy eigenvalues of the $^7$B resonances together with the many-body continuum cuts on the complex energy plane,
which is useful to understand the positions of poles and the various thresholds relatively at glance.
The $^6$Be resonances together with a last proton compose the thresholds of $^7$B,
whose positions are located at the starting points of the $-2\theta$-rotated cuts in CSM.
The energy of the $^7$B ground state is obtained as $E_r$=3.35 MeV and agrees with the recent experiment of $E_r=3.38(3)$ MeV\cite{charity11}.
The decay width is 0.49 MeV, which is good but slightly smaller than the experimental value of 0.80(2) MeV.
In the experiment, the decay width is determined from the $R$-matrix theory on the assumption 
of the decay into the $^6$Be($0^+_1$)+$p$ channel. On the other hand, our analysis shows that the $^6$Be($2^+_1$)-$p$ component is important in the $^7$B ground state, 
which is found from the $S$-factors of this channel and is suggested from the conventional shell model calculation\cite{charity11}.
There is no experimental evidence for the excited states of $^7$B so far and it is desired that further experimental data are coming.

\begin{table}[t]
\caption{Energy eigenvalues of the $^7$B resonances measured from the $^4$He+3$p$ threshold.
The values with parentheses are the experimental ones\cite{charity11}. 
Dominant configurations are listed.}
\label{ene7}
\centering
\begin{ruledtabular}
\begin{tabular}{c|cccc}
            & Energy~[MeV]     & Width~[MeV]    & Configuration          \\ \hline
 $3/2^-_1$  & $3.35$~(3.38(3)) & 0.49~(0.80(2)) & $(p_{3/2})^3$          \\
 $3/2^-_2$  & $6.92$           & 5.422          & $(p_{3/2})^2(p_{1/2})$ \\
 $3/2^-_3$  & $8.39$           & 9.86           & $(p_{3/2})(p_{1/2})^2$ \\
 $1/2^-  $  & $5.93$           & 4.73           & $(p_{3/2})^2(p_{1/2})$ \\
 $5/2^-  $  & $4.63$           & 3.91           & $(p_{3/2})^2(p_{1/2})$ \\
\end{tabular}
\end{ruledtabular}
\end{table}

\begin{table*}[ht]
\centering
\caption{Dominant configurations of three valence protons in the $^7$B resonances with their squared amplitudes $(C^J_c)^2$.}
\label{conf}
\begin{ruledtabular}
\begin{tabular}{ll|ll|lrc}
\multicolumn{2}{c|}{$3/2^-_1$}          & \multicolumn{2}{c|}{$3/2^-_2$}          & \multicolumn{2}{c}{$3/2^-_3$}            \\ \hline 
$(p_{3/2})^3$          &$0.923+i0.002$  & $(p_{3/2})^2(p_{1/2})$  &$0.795+i0.032$ & $(p_{3/2})(p_{1/2})^2$  & $0.770+i0.053$ \\
$(p_{3/2})(p_{1/2})^2$ &$0.020+i0.004$  & $(p_{3/2})(p_{1/2})^2$  &$0.195-i0.035$ & $(p_{3/2})^2(p_{1/2})$  & $0.182-i0.050$ \\
$(p_{3/2})^2(p_{1/2})$ &$0.021-i0.007$  & $(d_{3/2})^2(p_{3/2})$  &$0.006+i0.001$ & $(p_{3/2})^3$           & $0.003-i0.002$ \\
\end{tabular}
\vspace*{0.2cm}
\begin{tabular}{lr|lrc}
\multicolumn{2}{c|}{$1/2^-$}             & \multicolumn{2}{c}{$5/2^-$}                       \\ \hline	
$(p_{3/2})^2(p_{1/2})$  & $0.969-i0.000$ &  $(p_{3/2})^2(p_{1/2})$        & $0.957+i0.006$   \\	
$(d_{5/2})^2(p_{1/2})$  & $0.018-i0.002$ &  $(d_{3/2})(d_{5/2})(p_{3/2})$ & $0.015-i0.003$   \\	
$(1s_{1/2})^2(p_{1/2})$ & $0.005+i0.002$ &  $(d_{3/2})^2(p_{3/2})$        & $0.008-i0.001$   \\	
\end{tabular}
\end{ruledtabular}
\end{table*}

\begin{figure}[t]
\centering
\includegraphics[width=8.5cm,clip]{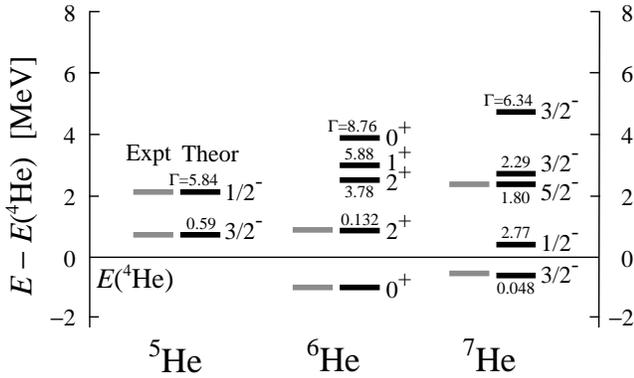}
\caption{Energy levels of He isotopes measured from the $^4$He energy. Units are in MeV.
Black and gray lines are theory and experiments, respectively. Small numbers are decay widths.}
\label{fig:He7}
\end{figure}

\begin{figure}[th]
\centering
\includegraphics[width=8.0cm,clip]{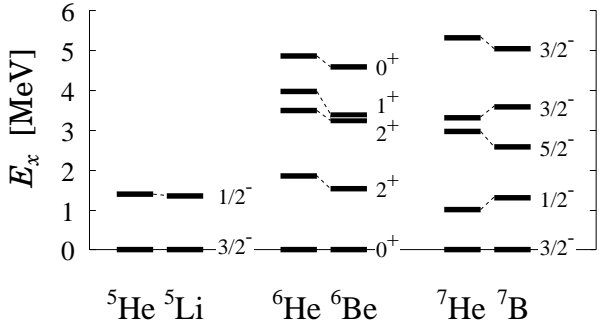}
\caption{Excitation energy spectra of mirror nuclei of A=5,6,7 in the units of MeV.}
\label{fig:excite}
\end{figure}

We discuss the configuration properties of each resonance of $^7$B in detail.
In Table \ref{conf}, we list the main configurations with their squared amplitudes $(C^J_c)^2$ in Eq. (\ref{WF0}) for each $^7$B resonance.   
In general, the squared amplitude of resonant state can be a complex number, while the total of the squared amplitudes is normalized as unity. The interpretation of the imaginary part in the physical quantity of resonances is still an open problem\cite{homma97}. 
In the results of $^7$B, the amplitudes of the dominant components are almost real values.
It is, hence, expected to discuss the physical meaning of the dominant components of the resonances in the same way as the bound state.
It is furthermore found that the imaginary parts of the configurations are canceled to each other for every resonance 
and their summations have much smaller imaginary parts.
When we consider all the available configurations, the summations conserve unity due to the normalization of the states.

For the $3/2^-$ ground state, 
the result indicates that the $(p_{3/2})^3$ configuration is dominant with a small mixing of the $p_{1/2}$ component.
For the excited $3/2^-_2$ state, one proton occupies the $p_{1/2}$ orbit and the residual two protons in $p_{3/2}$ form the spin of $2^+$, 
which corresponds to the $^6$Be($2^+_1$) configuration as shown in Table~\ref{comp6_2}.
The importance of the $^6$Be($2^+_1$)-$p$ component in the $3/2^-_2$ state of $^7$B is discussed from the viewpoint of the $S$-factors.
It is also found that the two-particle excitation into the $(p_{1/2})^2$ configuration is mixed by about 20\%.
The $3/2^-_3$ state is dominated by the $(p_{3/2})(p_{1/2})^2$ configuration,
in which the $(p_{1/2})^2$ part is the same configuration of $^6$Be($0^+_2$).

The $1/2^-$ state of $^7$B corresponds to the one particle excitation from the ground state.
Its decay width, 4.73 MeV is large and comparable to the resonance energy, 5.93 MeV among the five resonances of $^7$Be. 
This is confirmed from Fig. \ref{fig:ene_7B} as the large ratio of the imaginary part to the real one in the complex energy plane.
The result of the large decay width is similar to the $^5$Li($1/2^-$) state in the $^4$He+$p$ system.
In comparison with $^5$Li, whose resonance energy is 2.93 MeV with the decay width of 6.49 MeV,
the $^7$B($1/2^-$) state of has a smaller decay width.
This difference comes from the residual two protons occupying the $p_{3/2}$ orbit in $^7$B.
The attractive contribution between the $p_{1/2}$ proton and other two protons makes the decay width of the $1/2^-$ state smaller.
In the $5/2^-$ state, the $2^+$ component of $(p_{3/2})^2$ plus $p_{1/2}$ is dominant.
This coupling scheme is similar to the $3/2^-_2$ case.
In relation to the configuration properties of $^7$B, it is interesting to examine the $^6$Be-$p$ components in each $^7$B state, which is performed using the $S$-factors.

It is interesting to discuss the mirror symmetry between $^7$B and $^7$He consisting of $^4$He and three valence protons or neutrons.
To do this, we show the energy spectra of He isotopes with COSM in Fig.~\ref{fig:He7}, 
using the Hamiltonian in Eq.~(\ref{eq:Ham}) without the Coulomb term. 
The experimental data of $^7$He($1/2^-$) is not fixed\cite{Me02,Bo05,Sk06,Ry06,beck07,Wu08}, so that we do not put the data in the figure.
From Figs.~\ref{fig:B7} and \ref{fig:He7}, it is found that the order of energy levels are the same between proton-rich and neutron-rich sides.
In the proton-rich side, the whole spectra are shifted up due to the Coulomb repulsion 
in comparison with those of the neutron-rich side.
The displacement energies are about 2.5 MeV for $^6$Be from $^6$He,
and about 4 MeV for $^7$B from $^7$He, respectively.
In Fig. \ref{fig:excite}, we compare the excitation energy spectra of proton-rich and neutron-rich sides.
It is found that the good symmetry is confirmed between the corresponding nuclei.
The differences of excitation energies for individual levels are less than 1 MeV.
The properties of the configurations of $^7$B and $^7$He are discussed in terms of $S$-factors, next.

\subsection{Spectroscopic factors of $^7$B}\label{sec:sfac}

We obtain the information of the structures of $^7$B via the $S$-factors.  
In this study, we extract the $S$-factors of the $^6$Be-$p$ components in $^7$B.
This quantity is important to examine the coupling behavior between $^6$Be and a last proton
including the excitations of $^6$Be.
We choose the $0^+_1$ and $2^+_1$ states of $^6$Be, which are observed experimentally.
In this analysis, both of initial ($^7$B) and final ($^6$Be) states are resonances, so that the $S$-factors become complex numbers.
The present $S$-factors correspond to the components of $^6$Be in the $^7$B resonances 
and contain the imaginary parts.
It is still difficult to derive the definite conclusion of the interpretation of the imaginary part in the $S$-factors, as was mentioned in the previous studies \cite{myo09}.
The further theoretical and mathematical developments would be desired to solve this problem.

\begin{figure}[t]
\centering
\includegraphics[width=7.5cm,clip]{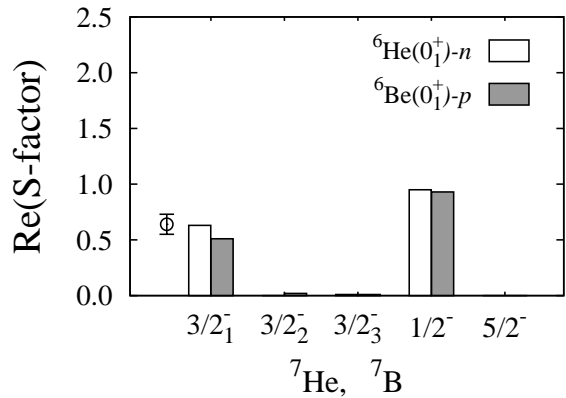}
\caption{Real part of the $S$-factors of $^7$B and $^7$He, in which the daughter nuclei are the $0^+_1$ states.
The experimental data of the $^7$He($3/2^-$) state\cite{beck07} is shown by the open circle.}
\label{fig:sfac0}
\end{figure}

\begin{figure}[t]
\centering
\includegraphics[width=7.5cm,clip]{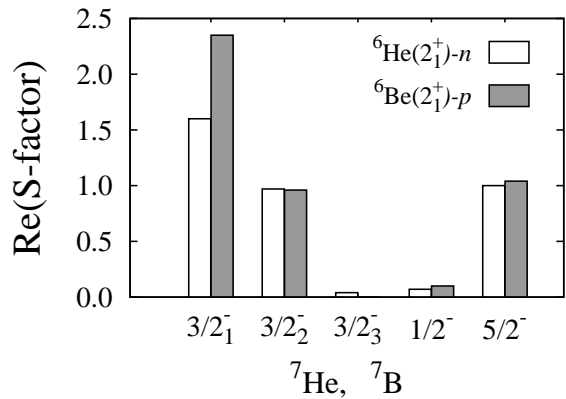}
\caption{Real part of the $S$-factors of $^7$B and $^7$He, in which the daughter nuclei are the $2^+_1$ states.}
\label{fig:sfac2}
\end{figure}

In Table~\ref{sf_B7}, we list the results of $S$-factors of $^7$B.
For comparison, the results of $^7$He are shown in Table~\ref{sf_He7}.
It is found that most of the components show almost the real values in $^7$B and $^7$He.
Hence, the comparison of the real parts of the $S$-factors for $^7$B and $^7$He is shown in Figs.~\ref{fig:sfac0} and \ref{fig:sfac2}.

\begin{table}[t]
\caption{$S$-factors of the $^6$Be-$p$ components in $^7$B. Details are described in the text.}
\label{sf_B7}
\begin{ruledtabular}
\begin{tabular}{c|ccc}
          & $^6$Be($0^+_1$)-$p$ &  $^6$Be($2^+_1$)-$p$ \\ \hline
$3/2^-_1$ & $0.51+i0.02$        &  $2.35-i0.15$   \\ 
$3/2^-_2$ & $0.02-i0.01$        &  $0.96-i0.01$   \\
$3/2^-_3$ & $0.00+i0.01$        & $-0.01-i0.06$   \\ 
$1/2^- $  & $0.93-i0.02$        &  $0.10-i0.01$   \\
$5/2^- $  & $0.00+i0.00$        &  $1.04-i0.01$   \\
\end{tabular}
\end{ruledtabular}
\end{table}
\begin{table}[t]
\caption{$S$-factors of the $^6$He-$n$ components in $^7$He. Details are described in the text.}
\label{sf_He7}
\begin{ruledtabular}
\begin{tabular}{c|ccc}
          & $^6$He($0^+_1$)-$n$ &  $^6$He($2^+_1$)-$n$ \\ \hline
$3/2^-_1$ & $0.63+i0.08$        &  $1.60-i0.49$   \\
$3/2^-_2$ & $0.00-i0.01$        &  $0.97+i0.01$   \\
$3/2^-_3$ & $0.01+i0.00$        &  $0.04-i0.01$   \\ 
$1/2^- $  & $0.95+i0.03$        &  $0.07-i0.02$   \\
$5/2^- $  & $0.00+i0.00$        &  $1.00+i0.01$   \\
\end{tabular}
\end{ruledtabular}
\end{table}


In Table \ref{sf_B7}, for the $3/2^-_1$ state, the $^6$Be($2^+_1$)-$p$ component is large, more than four times of that of the $^6$Be($0^+_1$)-$p$ component for real part.
This means that the $^6$Be($2^+_1$) state is dominant in this state.
The similar trend can be seen in $^7$He in Table \ref{sf_He7}, where the real part of the $^6$He($0^+_1$)-$n$ component
agrees with the observation of 0.64(9) \cite{beck07}, as shown in Fig.~\ref{fig:sfac0}.
For the $3/2^-_2$ state, the $^6$Be($2^+_1$)-$p$ component is selectively mixed from the dominant amplitude of $(p_{3/2})^2_{2^+}\otimes(p_{1/2})$.
For the $3/2^-_3$ state, the $0^+_1$ and $2^+_1$ states of $^6$Be are hardly included because of the $(p_{3/2})\otimes(p_{1/2})^2$ configuration.
Instead of the above two $^6$Be states, the $^6$Be($0^+_2$) state with $(p_{1/2})^2$ configuration and the $^6$Be($2^+_2$) state with $(p_{3/2})(p_{1/2})$ one 
may give large contributions for this state.
For the $1/2^-$ state, the $S$-factor of $^6$Be($0^+_1$)-$p_{1/2}$ proton is close to unity with a small imaginary part and the $^6$Be($2^+_1$)-$p$ component is small.
Hence, the $^6$Be($0^+_1$)-$p$ component is dominant in the $1/2^-$ state. 
The large mixing of the $0^+$ state of $A=6$ nuclei is also confirmed in the $^7$He($1/2^-$) state as shown in Table \ref{sf_He7}.
In $^7$He($1/2^-$), we have suggested the weak coupling nature of the $p_{1/2}$ orbital neutron around $^6$He, which retains a two-neutron halo structure\cite{myo09}.
For the $5/2^-$ state of $^7$B, the $^6$Be($2^+_1$)-$p$ component is included well, similar to $3/2^-_2$ as was explained.
These two states have a similar structure of the configurations of valence protons.
From the $S$-factor analysis, the most of the $^7$B states are not considered 
to be purely single particle states coupled with the $^6$Be ground state except for the $1/2^-$ state.
The component of $^6$Be($2^+_1$) is important in several states. This conclusion is the same as that of $^7$He.

We consider the structure differences between $^7$B and $^7$He from the $S$-factors and discuss the mirror symmetry.
From Fig.~\ref{fig:sfac2}, the sizable difference between the components including the $A=6$($2^+$) states is seen in the ground states of $^7$B and $^7$He.
The $^6$Be($2^+_1$)-$p$ component in $^7$B obtained as $2.35$ is larger than the $^6$He($2^+_1$)-$n$ component in $^7$He as $1.60$ by 47\% for real part.
The other four excited states show the similar values between two nuclei in Figs.~\ref{fig:sfac0} and \ref{fig:sfac2}.
In those excited states, either of the components of $0^+$ and $2^+$ of $A=6$ nuclei is selectively mixed.
These results indicate that the breaking of the mirror symmetry is occurred only in their ground states.
The reason of the difference in the $2^+$ coupling is that the $^7$B ground state is located closely 
to the $^6$Be($2^+_1$) state by 0.45 MeV for resonance energy, as shown in Fig.~\ref{fig:B7},
where the decay widths of two states are rather small in comparison with other resonances.
This situation is not occurred in $^7$He as shown in Fig.~\ref{fig:He7}, in which the energy difference between $^7$He($3/2^-_1$) and $^6$He($2^+_1$) is 1.46 MeV.
The small energy difference between $^7$B and $^6$Be($2^+_1$) enhances the $^6$Be($2^+_1$)-$p$ component in $^7$B as the coupling to the open channel of the $^6$Be($2^+_1$)+$p$ threshold.
On the other hand, the $^6$Be($0^+_1$)-$p$ component in $^7$B becomes smaller than that of $^7$He by 24 \% as shown in Fig.~\ref{fig:sfac0},
because the energy difference between the ground states of $^7$B and $^6$Be is 1.97 MeV, larger than the case of $^7$He of 0.40 MeV.
The origin of the difference of the $S$-factors in $^7$B and $^7$He is the Coulomb repulsion, which acts to shift the entire energies of the $^7$B states up.
The well-known effect of the Coulomb interaction to break the mirror symmetry is the Thomas-Erhman shift,
in which the $s$-wave dominant states suffer the different effect of Coulomb repulsion from the states having mainly other partial waves.
On the other hand, the present result found in the $^7$B ground state is caused by the existence of the several open channels including the excitations of subsystems and is different from the Thomas-Erhman shift.

As conclusion, the mirror symmetry is broken only in the ground states of $^7$B and $^7$He, while the excited states of two nuclei keep the symmetry. 
This result is associated with the energies of the $A=6$ subsystem as the open channels of the one nucleon emission.
It is experimentally desired to observe the $2^+$ components of $A=6$ nuclei in $^7$B and $^7$He and examine the mirror symmetry.
In the present analysis, the $S$-factors represent the contributions of only the resonances of $^7$B and $^6$Be. 
By considering the additional contributions of the remaining continuum states of two nuclei,
it is available to obtain the strength functions of the one-proton removal from $^7$B into $^6$Be and also into the $^4$He+$p$+$p$ final states, which are observable.
It is interesting to obtain these strengths and compare them with the one-neutron removal strength from $^7$He into $^6$He \cite{myo09}.

\section{Summary}\label{sec:summary}
We have investigated the resonance structures of $^7$B with the $^4$He+$3p$ four-body cluster model.
The boundary condition for many-body resonances is accurately treated using the complex scaling method. 
The decay thresholds concerned with subsystems are described consistently.
We have found five resonances of $^7$B, which are dominantly described by the $p$-shell configurations. 
The energy and the decay width of the ground state agree with the recent experiment.
We also predict four excited resonances of $^7$B, which are desired to be confirmed experimentally.

We further investigate the spectroscopic factors of the $^6$Be-$p$ components in $^7$B to examine the coupling behavior between $^6$Be and a last proton.
It is found that the $^6$Be($2^+_1$) state contributes largely in the ground and the several excited states of $^7$B.
In comparison with $^7$He, the mirror nucleus of $^7$B, the $^6$Be($2^+_1$)-$p$ component in the $^7$B ground state is larger than the $^6$He($2^+_1$)-$n$ component in the $^7$He ground state.
This difference comes from the fact that the $^7$B ground state is close to the $^6$Be($2^+_1$) state in energy by the Coulomb repulsion.
This situation enhances the $^6$Be($2^+_1$)-$p$ component in $^7$B as the channel coupling.
The different coupling of $A=6$ nuclei in $^7$B and $^7$He is occurred only in their ground states and indicates the breaking of the mirror symmetry.
It is desired to observe the difference of the couplings in $^7$B and $^7$He experimentally.

\section*{Acknowledgments}
We thank Professor Kiyomi Ikeda for fruitful discussions.
This work was supported by a Grant-in-Aid for Young Scientists from the Japan Society for the Promotion of Science (No. 21740194).

\section*{References}
\def\JL#1#2#3#4{ {{\rm #1}} \textbf{#2}, #4 (#3)}  
\nc{\PR}[3]     {\JL{Phys. Rev.}{#1}{#2}{#3}}
\nc{\PRC}[3]    {\JL{Phys. Rev.~C}{#1}{#2}{#3}}
\nc{\PRA}[3]    {\JL{Phys. Rev.~A}{#1}{#2}{#3}}
\nc{\PRL}[3]    {\JL{Phys. Rev. Lett.}{#1}{#2}{#3}}
\nc{\NP}[3]     {\JL{Nucl. Phys.}{#1}{#2}{#3}}
\nc{\NPA}[3]    {\JL{Nucl. Phys.}{A#1}{#2}{#3}}
\nc{\PL}[3]     {\JL{Phys. Lett.}{#1}{#2}{#3}}
\nc{\PLB}[3]    {\JL{Phys. Lett.~B}{#1}{#2}{#3}}
\nc{\PTP}[3]    {\JL{Prog. Theor. Phys.}{#1}{#2}{#3}}
\nc{\PTPS}[3]   {\JL{Prog. Theor. Phys. Suppl.}{#1}{#2}{#3}}
\nc{\PRep}[3]   {\JL{Phys. Rep.}{#1}{#2}{#3}}
\nc{\JP}[3]     {\JL{J. of Phys.}{#1}{#2}{#3}}
\nc{\andvol}[3] {{\it ibid.}\JL{}{#1}{#2}{#3}}

\end{document}